# Enhanced granular medium-based tube press hardening

Hui Chen[a], Sigrid Hess[a], Jan Haeberle[b], Sebastian Pitikaris[b], Philip Born[b], Alper Güner[a, 1], Matthias Sperl[b], A. Erman Tekkaya (1)[a, *]

[a]*Institute of Forming Technology and Lightweight Construction (IUL), TU Dortmund University, 44227 Dortmund, Germany*
[b]*Institute of Materials Physics in Space, German Aerospace Center (DLR), 51170 Köln, Germany*

Active and passive control strategies of internal pressure for hot forming of tubes and profiles with granular media are described. Force transmission and plastic deformation of granular medium is experimentally investigated. Friction between tube, granular medium and die as also the external stress field are shown to be essential for the process understanding. Wrinkling, thinning and insufficient forming of the tube establishes the process window for the active pressure process. By improving the punch geometry and controlling tribological conditions, the process limits are extended. Examples for the passive pressure process reveal new opportunities for hot forming of tubes and profiles.

Metal forming, hot stamping, hydroforming

## 1. Introduction

Press hardening of sheets, which combines hot forming and quenching, enables the production of components with strengths of over 1500 MPa, [1]. Hot forming of boron micro-alloyed steel sheets (basically 22MnB5) is state of the art, [2]. Crash-relevant components, for instance, also need high stiffness besides high strength. Closed profiles provide high structural rigidity at compact space. For forming hollow structural parts like sub-frames or axle components, tube hydroforming is widely used, [3]. High strength profiles can be manufactured by combining tube hydroforming and press hardening. This requires an adequate working medium withstanding the austenitization temperature of steel at 950°C to 1000°C. Gaseous fluid medium offers the potential for hot tube hydroforming, for instance of aluminium and titanium structures, [4]. Hydroforming of tubes made of 1.4512 ferritic stainless steel and press hardening steel 22MnB5 have been realised using gas at temperatures of 950°C and 1000°C respectively, [5]. A gas pressure generator has been used providing internal pressures of up to 80 MPa for tube press hardening, [6]. Gas is also used as passive forming medium for press hardening of torsion beams, [7]. In this application the heated tube is sealed on both sides and internal pressure is built up by compressing the tube. The critical cooling rate for martensitic transformation is achieved by applying compressed air. One disadvantage of using gaseous medium is that it only contributes to the cooling process by 15 % compared to the total tooling, [8]. The final hardness of tubes formed by gas differs locally and depends mainly on the pressure rate, the maximum pressure value, and the tool temperature, [9]. Another disadvantage of using gas is its high compressibility.

An alternative working medium for high temperature applications with high-pressure conditions is granular material, such as zirconia beads or quartz sand. Ceramic beads were used for sheet hydroforming of DC04 at 600°C, [10]. With granular medium, high temperatures of up to 1000 °C and high pressures of up to 100 MPa have been realised for press hardening of tubes using the granular material as active pressure medium, [11] (Fig. 1a). In this study it has been shown that granular material with less compressibility, lower internal friction, and optimum particle size improves the forming pressure. The Drucker–Prager Cap (DPC) plasticity model has proven to be suitable for thermal-mechanically coupled numerical modelling of the press hardening process. In [12] the idea of using a granular material as passive pressure medium (Fig. 1b) has been introduced for the first time.

This paper aims at utilising the physics of granular medium during hot forming to predict the processing window of press hardening of profiles with active and passive granular medium and at extending the limits of the process.

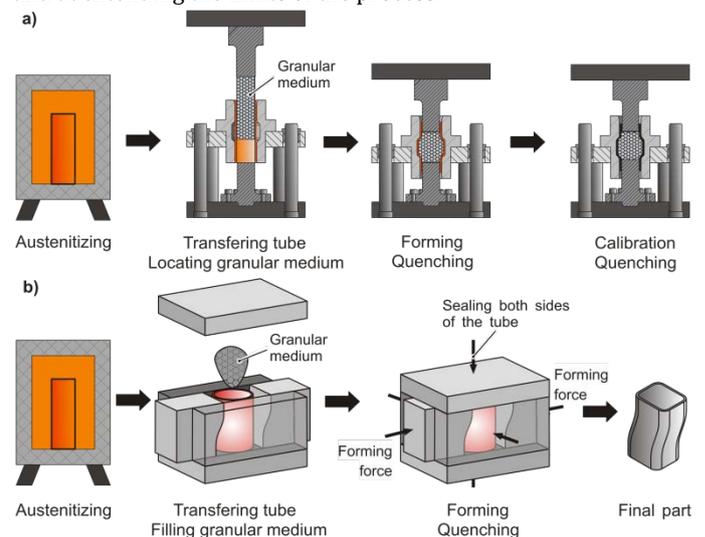

**Fig. 1.** Tube press hardening by granular medium: (a) active, (b) passive.

## 2. Granular material as forming medium

Force transmission and plastic deformation proceed fundamentally different in granular medium than in gas and liquid. Two-dimensional horizontal experiments were carried out to understand the behaviour of granular medium during the tube forming process. The process is represented by a die with rigid side walls confining the granular medium (epoxy resin discs, Vishay PS 4) through which a displacement-controlled punch is driven (Fig. 2a). Deformation of the medium can be monitored by tracking the disc positions, while internal forces are visualised by using stress-birefringence [13].

* Corresponding author.
  *E-mail address*: erman.tekkaya@iul.tu-dortmund.de (A.E. Tekkaya).
[1] Present address: AutoForm Engineering Deutschland GmbH, 44227 Dortmund, Germany.

*2.1 Force transmission in granular medium*

The pressure induced by a punch pushed into the granular medium provokes a ramified network of force chains, which involves only a fraction of all particles (Fig. 2a). When the granular medium is deformed the force chains form and break, leading to large fluctuations of the contact forces in space and time [14]. Thus, granular medium can be treated as homogeneous medium only on scales much larger than the particles.

The force chains are supported by individual particle contacts at the die walls and fade with growing distance to the punch. This behaviour induces a strong dependence of the force transmission on the friction between particles and die walls. Much stronger forces build up close to the punch when the die walls prevent slipping of the particles ('sticky boundary') than in a frictionless case ('slip boundary') for same punch displacement. These forces also fade at a stronger rate than in a die with discs free to slip at the die walls. A quantitative evaluation reveals that the decay of forces fits well with an exponential decay (Fig. 2b).

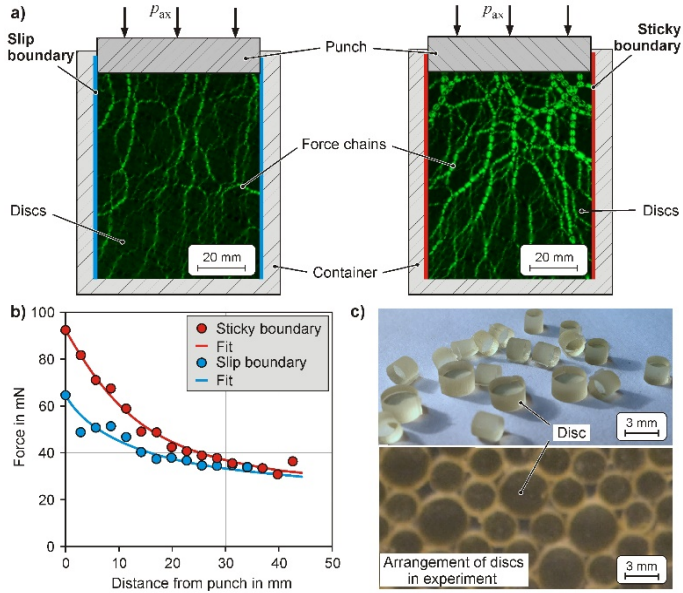

**Fig. 2.** Analysis of forces within granular medium in a compression test: a) visualisation of the force transmission for two friction levels; bright areas indicate particles under stress, b) total force with an exponential fit, c) geometry of the utilised discs.

*2.2 Plastic deformation of granular medium*

The granular medium deforms elasto-plastically, in contrast to gas and liquid usually used in tube hydroforming. The forming of a tube intrinsically requires the plastic deformation of the granular medium used. The deformation of granular medium occurs markedly localised, comparable with the formation of shear bands. The localisation is determined by the internal friction angle and the external stress field [15]. The latter makes the plastic deformation punch shape-dependent. The force chains align locally with the punch surface and spread more isotropically with inclined punch surfaces (Fig. 3a). This translates into different deformation behaviour of the granular medium in the compression experiments, as exposed by removing one side wall (Fig. 3b). Particle displacement is localised under compression by the flat punch, while the motion of the round punch induces a displacement nearly throughout the sample for the same punch displacement.

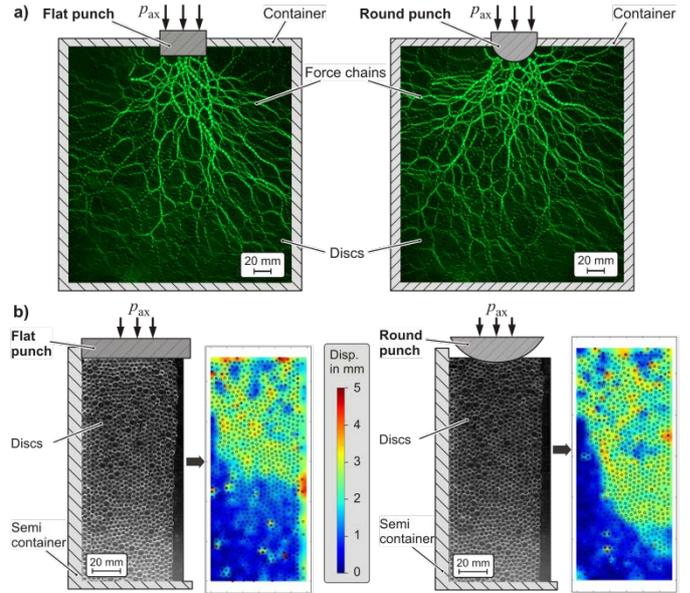

**Fig. 3.** Dependence of the force chain pattern (a) and plastic deformation of the granular medium (b) on the punch shape represented by the particle displacement. All figures are rotated by 90°.

In conclusion, the frictional nature of the granular medium and the localisation of plastic deformation can be expected to strongly affect the tube forming process. The friction among the particles and with the tube walls yields an exponential decay of externally applied forces within the medium, while the shape of the punch can enhance the force propagation and the plastic behaviour of the granular medium. This knowledge is used in the next section for understanding and extending the process window.

## 3. Enhancing active medium-based tube press hardening

The enhancement of active granular medium forming utilising the knowledge of the previous section is done by experiments and numerical analyses based on the generic tube expansion process specified in Fig. 4. The numerical (finite element) model is described in [11]. Zirconia beads and quartz sand are used as forming media. Only results for zirconia beads will be presented. The expanded outer radius of the tube $r_f$ is determined by the forming limit diagram of the workpiece material 22MnB5 given in [16] for the initial tube radius $r_0$ as the maximum possible deformed radius. In classical tube hydroforming external axial feeding is necessary to compensate tube thinning and to reduce the forming pressure. One inherent benefit of tube forming using granular medium is that the complicated external axial feeding can be replaced by the feeding through the frictional forces $f_g$ between the tube inner wall and the granular medium. The maximum shoulder angle $\alpha$ of the part is determined by the consolidation pressure of the granular medium, which is 220 MPa for zirconia beads. For the given dimensions this angle is 60°.

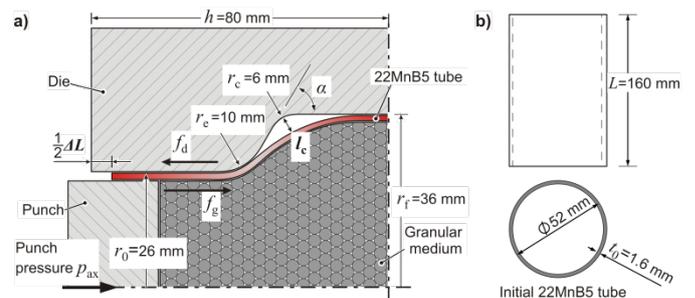

**Fig. 4.** a) Generic tool design for active granular medium-based hot forming of tubes, b) initial tube dimensions made of 22MnB5.

## 3.1. Enhancement by control of friction

Friction forces in granular medium-based tube hydroforming are crucial for the process success. There are two kinds of frictional forces with opposite senses: the internal friction force $f_g$ between the tube's inner surface and the granular medium, and the external friction force $f_d$ between the tube's external surface and the die (Fig. 4). $\mu_d$ and $\mu_g$ are the respective Coulomb friction coefficients. The friction force $f_g$ acts as axial feeding force. It is beneficial to reduce the tube thinning. But it also reduces the radial pressure with increasing axial position as shown in Section 2.1 (Fig. 2). On the other hand, too high feed forces might cause buckling of the tube in the bulge area. The friction force $f_d$ hinders the axial material flow and acts against the feeding force $f_g$.

By using dry or lubricated conditions (with cubic boron nitride (CBN) solid lubricant) at the tube inner and outer surface, the effect of the frictional forces is investigated. The friction coefficients are determined by the direct shear test according to [11]. Higher coefficients of frictions (Fig. 5) apply to the dry state.

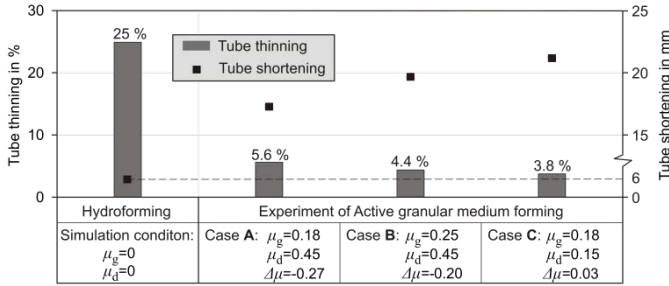

**Fig. 5.** Experiment of active granular medium forming: thinning and shortening of tube under different tribological conditions.

Tube thinning is drastically reduced by the active granular medium forming compared to conventional tube hydroforming without axial feeding (Fig. 5). The positive feed force (= $f_g$ - $f_d$) reduces wall thinning (Fig. 5) and, if excessive, promotes buckling (wrinkling) in the bulge area (Fig. 6). A high friction force $f_g$ reduces the bulging pressure and may lead to incomplete forming (Case C). The results for quartz sand (not shown) have the same tendencies. The conclusion of the analysis of frictional forces indicates that the best condition (low thinning, no buckling and complete forming) is given for Case B with moderate $\Delta\mu$.

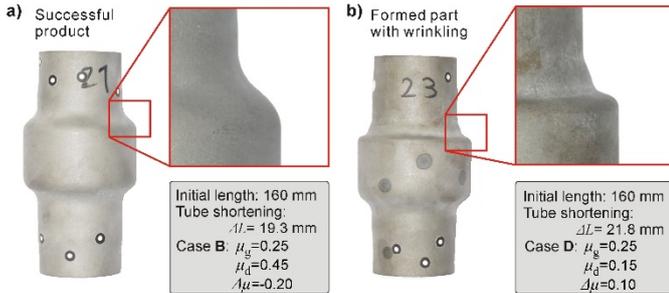

**Fig. 6.** Effect of tribology: a) Successful part (low thinning, no buckling and complete forming), b) wrinkling due to excessive feeding

## 3.2. Enhancement by formation of multiple shear bands

A complex shear state can enhance the plasticity of the granular medium, as shown in Section 2.2 (Fig. 3). The multiple shear band formation leads to a better radial pressure performance of the granular medium. Using this insight, four different punch shapes have been investigated experimentally and numerically to form bulges in tubes with various shoulder angles $\alpha$. Fig. 7 shows the essential, i.e. the minimum, punch pressures to fully form the tubes. The shape is assumed to be fully formed if the clearance between the tube and the die corner is smaller than the tube's initial thickness. This is also essential to ensure the rapid cooling necessary for the martensitic formation, as extensive hardness measurements have confirmed. As the shoulder gets steeper naturally higher radial pressure is required. At the limiting axial pressure of 220 MPa is zirconia beads are consolidated into green body and fail. Two experiments were carried out for the shoulder angle 60° using flat and conical punches for which the limiting pressure is not exceeded. All remaining data was obtained by numerical analyses. The highest punch pressure for a specific shoulder angle is given for the flat punch and the lowest one for the round punch. For the shoulder angle 75° the difference in pressure for various punch shapes is 80 MPa (40 %) so that a 75° shoulder angle can be formed only with a round punch if zirconia beads are used as the pressure medium.

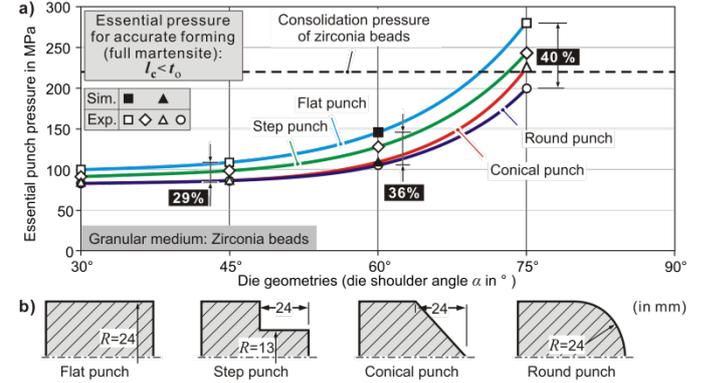

**Fig. 7.** a) Minimum punch pressure using different punch geometries for various forming geometries, b) analysed punch geometries.

## 3.3. Process window

Typical defects in active granular medium-based tube press hardening processes are insufficient forming, buckling, and excessive thinning. For flat and conical punches the process windows for the workpiece geometry given in Fig. 4a with 60° shoulder angle are shown in Fig. 8a. One natural process parameter is the axial punch pressure. The other process parameter is the difference in the friction coefficients $\Delta\mu = \mu_g - \mu_d$. Large values of $\Delta\mu$ indicate large feed forces. For a given axial punch pressure, buckling and thinning can be controlled by the tribological conditions at the tube-die and tube-medium interfaces, allowing the enhancement of the process. The corresponding effective feeding, normalised as the engineering axial strain in the bulging zone, is shown in Fig. 8b. The thinning limit is set to be 5 %. The influence of the punch shape on the process window is also shown. The conical punch allows higher formability and less buckling for the same punch pressure, whereas the thinning process limit becomes worse.

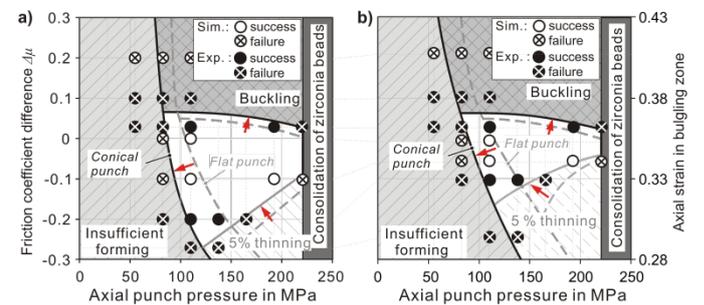

**Fig. 8.** Process windows of active granular medium-based tube press hardening (60° die shoulder angle, CBN as lubricant): a) using friction difference, b) using axial feeding strain.

## 4. Passive granular medium-based tube press hardening

The advantage of the passive granular medium-based tube pressing shown in Fig. 1b is that the pressure loss due to interfacial friction between the granular medium and the tube

(Section 2.1, Fig. 2b) is minimized since the external pressure is applied directly on the circumference of the profile cross-section. Additionally, the complex section geometry induces multiple shear bands (Section 2.2, Fig. 3), leading to homogeneous plastification of the internal granular medium. With these mechanisms the pressure generation by the granular medium can be achieved more efficiently than in active hydroforming of profiles. To validate these thoughts, several preliminary experiments have been conducted in which a tubular workpiece is formed into a curved dog-bone profile (Fig. 9). In the given example the circumference of the initial cross-section and the finally formed cross-section is identical, leading to a reduction of the cross-sectional area and, hence, to a volumetric compressive strain of $\varepsilon_{vol}$ of 58% in the medium during forming.

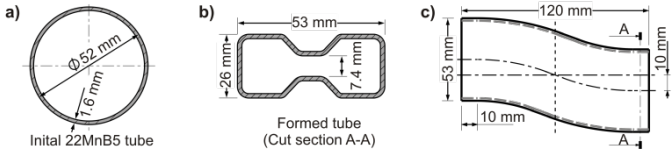

**Fig. 9.** Passive granular medium-based tube press hardening: a) initial cross-section, b) formed cross-section, c) top view of the component.

It is assumed that the forming of the bottom radius in passive forming can be modelled as the forming of a radius in hydroforming (Fig. 10). Based on the membrane theory, the needed hydrostatic pressure $p_h$ to form the radius from $r_0$ to $r$ is

$$p_h = \frac{2t_0 K}{\sqrt{3}r}\left(\left[\frac{2}{\sqrt{3}}\left(\frac{\pi}{4}-1\right)ln\left(\frac{r}{r_0}\right)\right]^n \left(\frac{r}{r_0}\right)^{\left(\frac{\pi}{4}-1\right)} + k_{f0}\right), \quad (3)$$

where the strength coefficient $K$, the strain hardening index $n$, and the initial yield stress $k_{f0}$ are material parameters of the flow curve extrapolation according to Ludwik [17].

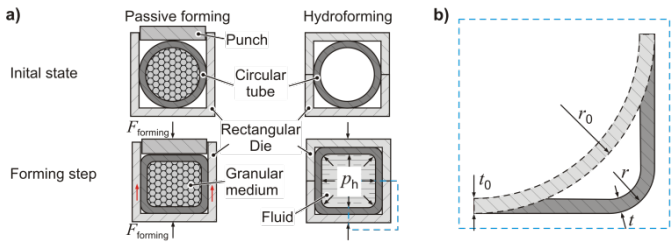

**Fig. 10.** a) Process principles of passive forming and hydroforming, b) incremental forming of corner from radius $r_0$ to $r$.

According to [11], the relation between volumetric strain and hydrostatic pressure of the granular medium (zirconia beads) under compression in passive forming is

$$\varepsilon_{vol} = 0.0634(1 - e^{-0.01274 p_h}), \ p_h \text{ in MPa}. \quad (4)$$

The volumetric strain of the granular medium can be controlled by filling the initial tube with different amounts of the medium. Analytically, a minimum filling ratio of 53 % for forming radii of 4 mm is needed according to Eqs. (3) and (4). The part could be formed successfully using granular medium compared to the hot formed part without passive working medium (Fig. 11). Folding and buckling are typical failure modes of this process (Fig. 11a).

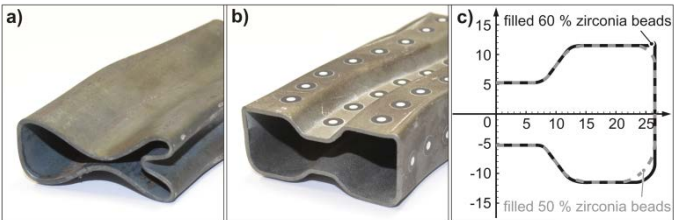

**Fig. 11.** (a) Hot-formed tube without working medium, (b) hot-formed part with passive granular medium (c) effect of volume filling amount on forming result.

## 5. Conclusions

The experimental analysis of the physics of forming a granular medium under high pressures revealed that

- Friction within the granular medium and with the tube walls exponentially reduces the radial pressure produced by the axial pressure with distance from the punch.
- Localisation of plastic deformation promoting the radial pressure effect of the medium depends strongly on the punch and the tube's cross-sectional geometries.

This knowledge has been used to improve the active granular medium-based hot forming of tubes by controlling the frictional conditions in the process and by modifying the punch geometry. For a given final tube geometry the process windows have been determined using physical process parameters of friction-difference between tube/medium and tube/die, or, the axial feeding strain and the axial punch pressure. The intrinsic axial feed generated by the granular medium has been identified as an advantage of the process.

Based on the physical insight of the granular medium under pressure, the potential of passive granular medium-forming of non-straight profiles has been demonstrated by preliminary experiments.

## Acknowledgement

This work is kindly supported by the German Research Foundation (DFG) under the grant numbers TE 508/52-1, BO 4174/2-1 and SP 714/12-1.

## References

[1] Neugebauer, R., Altan, T., Geiger, M., Kleiner, M., Sterzing, A., 2006, Sheet metal forming at elevated temperatures, Annals of the CIRP, 55/2: 793-816.
[2] Karbasian, H., Tekkaya, A. E., 2010, A review on hot stamping, J. Mater. Process. Technol., 210/15: 2103-2118.
[3] Koç, M., Altan, T., 2001. An overall review of the tube hydroforming (THF) technology, J. Mater. Process. Technol., 108/3: 384–393.
[4] Dykstra, B., Pfaffmann, G. D., Wu, X., 1999, Hot Metal Gas Forming – The next generation process for manufacturing vehicle structure components, International Body Engineering Conference, SAE Technical Paper Series, vol. 1999-01-3229.
[5] Vadillo, L., Santos, M. T., Gutierrez, M. A., Pérez, I., González, B., Uthaisangsuk, V., 2007, Simulation and experimental results of the hot metal gas forming technology for high strength steel and stainless steel tubes forming, AIP Conf. Proc. 908, 1199.
[6] Neugebauer, R., Schieck, F., 2010, Active media-based form hardening of tubes and profiles, Production Engineering, 4/4: 385-390.
[7] Maeno, T., Mori, K., Adachi, K., 2014, Gas forming of ultra-high strength steel hollow part using air filled into sealed tube and resistance heating, J. Mater. Process. Technol., 214/1: 97-105.
[8] Drossel, W., Pierschel, N., Paul, A., Katzfuß, K., Demuth, R., 2014, Determination of the active medium temperature in medium based press hardening processes, J. Manuf. Sci. Eng., 136/2, 021013-021013-8.
[9] Paul, A., Strano, M., 2016. The influence of process variables on the gas forming and press hardening of steel tubes, J. Mater. Process. Technol., 228: 160-169.
[10] Grüner, M., Gnibl, T., Merklein, M. 2014, Blank hydroforming using granular material as medium - investigations on leakage, Procedia Engineering, 81/11: 1035-1042.
[11] Chen, H., Güner, A., Ben Khalifa, N., Tekkaya, A. E., 2015, Granular media-based tube press hardening, J. Mater. Process. Technol., 228: 145-159.
[12] Chen, H., Hess, S., Güner, A., Tekkaya, A.E., 2015, Active and passive granular media-based tube press hardening, Tagungsband zum 10. Erlanger Workshop Warmblechumformung 2015, 115-129.
[13] Majmudar, T.S., Sperl, M., Luding, S., Behringer, R.P., 2007. Jamming transition in granular systems. Physical review letters, 98: 058001.
[14] Behringer, R.P., Daniels, K.E., Majmudar, T.S., Sperl, M., 2008, Fluctuations, correlations and transitions in granular materials: statistical mechanics for a non-conventional system, Philos. Trans. A. Math. Phys. Eng. Sci., 366: 493–504.
[15] Nedderman, R.M., 2005, Statics and kinematics of granular materials. Cambridge University Press.
[16] Bariani, P.F., Bruschi, S., Ghiotti, A., Turetta, A., 2008, Testing formability in the hot stamping of HSS, Annals of the CIRP, 57/1: 265-268.
[17] Marciniak, Z., Duncan, J.L., Hu, S.J., 2002, Mechanics of sheet metal forming, Butterworth-Heinemann, Oxford